\documentclass[12pt]{article}
\makeatletter
\@addtoreset{equation}{subsection}
\makeatother

\begin{document}
\begin{center}
{\Large \bf Cosmological Quantum Jump Dynamics} \\[0.5cm]
{\large\bf I. The Principle of Cosmic Energy Determinacy,\\
Equations of Motion, and Jump Probabilities }\\[1.5cm] {\bf
Vladimir S.~MASHKEVICH}\footnote {E-mail:
Vladimir\_Mashkevich@qc.edu}
\\[1.4cm] {\it Physics Department
  \\ Queens College\\ The City University of New York\\
  65-30 Kissena Boulevard\\ Flushing, New York
  11367-1519} \\[1.4cm] \vskip 1cm

{\large \bf Abstract}
\end{center}

The universe, as a closed system, is for all time in a state with
a determinate value of energy, i.e., in an eigenstate of the
Hamiltonian. That is the principle of cosmic energy determinacy.
The Hamiltonian depends on cosmic time through metric. Therefore
there are confluence and branch points of energy levels. At branch
points, quantum jumps must happen to prevent the violation of
energy determinacy. Thus quantum jumps are a reaction against the
propensity of the universe dynamics to that violation. On the
basis of this idea, an internally consistent quantum jump dynamics
is developed.

\newpage

\section*{Introduction}

Quantum jumps constitute the subject matter of the probabilistic
aspect of quantum physics. A standard method for attacking the
problem of quantum jumps has its basis in a unitary dynamics and
the concept of decoherence, i.e., a transformation of a pure state
into a mixed one. Such an approach can be traced back to the very
advent of quantum theory [1]. There exists a voluminous literature
on that approach. But it is impossible to construct an internally
consistent dynamics of quantum jumps on the aforementioned basis,
for unitary time evolution allows no decoherence.

In his lectures on gravitation [2], Feynman pointed out: ``It is
still possible that quantum theory does not absolutely guarantee
that gravity {\it has} to be quantized... If there were some
mechanism by which the phase evolution had a little bit of
smearing in it, so it was not absolutely precise, then our
amplitudes would become probabilities... But surely, if the phases
did have this built-in smearing, there might be some consequences
to be associated with this smearing. If one such consequence were
to be the existence of gravitation itself, then there would be no
quantum theory of gravitation... we should always keep in mind the
possibility that quantum mechanics may fail, since it has certain
difficulties with the philosophical prejudices that we have about
measurement and observation.''

Not quantizing gravity leads to semiclassical theory of
gravitation. In the conventional treatment of semiclassical
gravity, time evolution of a state vector $\Psi$  of the matter of
the universe is given by the Schr\"odinger equation
$d\Psi/dt=-{\rm i}H_t\Psi $, where the Hamiltonian $H_t$ depends
on time through metric $g$. This time evolution, however, is
unitary, which gives no way of constructing a consistent quantum
jump dynamics.

But Feynman's considerations perceived in a broad sense imply that
semiclassical gravity would be justified if it enabled us to
construct an internally consistent quantum jump dynamics.

Along these lines, crucial is the following observation:
Experience leads us to conclude that a closed system is for all
time in a stationary state, i.e., in a state with a determinate
value of energy. The universe is a closed system, so its energy
should have a determinate value. That is the principle of cosmic
energy determinacy. As the Hamiltonian $H_t$ is time-dependent,
the Schr\"odinger equation would have immediately violated the
principle and should be abandoned.

The Hamiltonian is $H_t=\sum_l\varepsilon_l(t)P_l(t)$ where $P_l$
is the projector for a level $l$. Denote a state vector by
$|\;\rangle\equiv\Psi$ and the corresponding state projector by
$P\equiv|\;\rangle\langle\:|$. Let the state belong to a level
$l$, so that $H_t\Psi_t=\varepsilon_l(t)\Psi_t$. In this case, the
state projector will be denoted by $P^l$. (Thus, subscript $l$
designates a level, and superscript $l$ a state belonging to that
level.) Now $H_tP^l(t)=P^l(t)H_t=\varepsilon_l(t)P^l(t)$. For a
degenerate level those conditions are insufficient. An equation of
motion for $P^l$ needs to be introduced. This is the one:
$dP^l/dt=(dP_l/dt)P^l+P^l(dP_l/dt)$.

As the Hamiltonian depends on time, there are confluence and
branch points of energy levels. At branch points, energy
determinacy would be violated. It is quantum jumps that prevent
the violation. Thus quantum jumps are a reaction against the
propensity of the universe dynamics to the violation of energy
determinacy.

As to the time $t$, it is universal cosmological time, the
existence of which is implied by quantum jumps themselves: they
give rise to a family of sets of simultaneous events---jumps of
$\ddot g\equiv\partial^2 g/\partial t^2$.

The approach outlined above has been advanced in [3]; some results
were obtained in [3-5]. The aim of this paper is to develop an
internally consistent quantum jump dynamics.

\section{The universe as a physical system}

\subsection{The Einstein equation and quantum jumps}

We adopt the classical description of spacetime and quantum
treatment of matter. The Einstein equation takes the form of

\begin{equation}
\label{1.1.1} G-\Lambda g=8\pi\kappa \tilde T
\end{equation}
\begin{equation}
\label{1.1.2} \tilde T=(\Psi,T\Psi)={\rm Tr}\{PT\}
\end{equation}
where $G$ is the Einstein tensor, $g$ is the metric, $\Lambda$ is
the cosmological constant, $\kappa$ is the gravitational constant,
$T$ is the energy-momentum tensor operator, $\Psi$ is a state
vector, and
\begin{equation}
\label{1.1.3} P=(\Psi,\cdot)\Psi=|\;\rangle\langle\,|\:\;,\quad
|\;\rangle =\Psi,\quad {\rm Tr}\,P=1
\end{equation}
is the corresponding state projector. We abandon the concept of
decoherence, so it is reasonable to exploit $P$ rather than
$\Psi$.

A jump of $P$ results in that of $\tilde{T}$. Write down
(\ref{1.1.1}) in components
\begin{equation}
\label{1.1.4} G_{\mu\nu}-\Lambda g_{\mu\nu}=8\pi\kappa\tilde
T_{\mu\nu},\quad \tilde T_{\mu\nu}={\rm
Tr}\{PT_{\mu\nu}\},\quad\mu,\nu=0,1,2,3
\end{equation}
The components $G_{ij}\: (i,j=1,2,3)$ of the Einstein tensor
involve the second time derivatives
\begin{equation}
\label{1.1.5} \ddot g_{ij}\equiv g_{ij,00}
\end{equation}
of the metric tensor components $g_{ij}$ [6,7]. That makes it
possible to retain the six equations
\begin{equation}
\label{1.1.6} G_{ij}-\Lambda g_{ij}=8\pi\kappa\tilde T_{ij}
\end{equation}
unchanged. Jumps of the $ \tilde T_{ij}$ will result in those of
the $\ddot g_{ij}$, which is quite conceivable from the physical
point of view: A jump of the force $\tilde T_{ij}$ results in a
jump of the acceleration $\ddot g_{ij}$. As to the four equations
\begin{equation}
\label{1.1.7} G_{0\mu}-\Lambda g_{0\mu}=8\pi\kappa\tilde{T}_{0\mu}
\end{equation}
the situation is completely different. The components $G_{0\mu}$
involve no second time derivatives of the $g_{\mu\nu}$; the only
time derivatives of metric involved in $G_{0\mu}$ are the $\dot
g_{ij}$. The latter should be continuous, not to mention the
$g_{\mu\nu}$. The violation of the four equations (\ref{1.1.7}) is
intolerable and they must be extended, which has been done in [5].

\subsection{Quantum-jump universal time}

A quantum jump of the state projector $P$ gives rise to a set of
events---jumps of $\ddot g$. These events are, by definition,
simultaneous, which allows for synchronizing clocks and thereby
furnishing the universal time. It is natural to identify
phenomenological cosmological time with the quantum-jump universal
time. Thus cosmological time is defined on the level of
fundamental physical laws---in contrast to the phenomenological
approach in classical cosmology [8,9].

In special relativity, the concept of simultaneity in connection
with quantum jumps makes no operationalistic sense, which
complicates the incorporation of quantum jumps into special
relativity. Taking gravity into account endows the concept with an
operationalistic content.

\subsection{Spacetime}

The universal cosmological time gives rise to a family of sets of
simultaneous events, thereby endowing spacetime with a fiber
structure. The metric compatible with this structure, i.e.,
admitting the synchronization of clocks, is of the form [7]
\begin{equation}
\label{1.3.1} ds^2=g_{00}(dx^0)^2+g_{ij}dx^idx^j
\end{equation}
or with
\begin{equation}
\label{1.3.2} dt=\sqrt{g_{00}}dx^0,\quad t=t(x^0,\vec x)=\int
g_{00}^{1/2}(x^0,\vec x)dx^0,\quad \vec x=(x^1,x^2,x^3)
\end{equation}
\begin{equation}
\label{1.3.3} ds^2=dt^2+g_{ij}dx^idx^j,\quad g_{ij}=g_{ij}(t,\vec
x)
\end{equation}
which relates to a synchronous reference frame. The latter, in its
turn, implies the product spacetime manifold
\begin{equation}
\label{1.3.4}
M=M^4=T\times S,\quad M\ni p=(t,s),\quad t\in
T,\quad s\in S
\end{equation}
The one-dimensional manifold $T$ is the universal cosmological
time, and the three-dimensional manifold $S$ is a cosmological
space. By (\ref{1.3.4}), the tangent space $M_p$ at a point $p\in
M$ is
\begin{equation}
\label{1.3.5} M_p=T_t\oplus S_s,\quad p=(t,s)
\end{equation}
and, in view of (\ref{1.3.3}),
\begin{equation}
\label{1.3.6} T_t\;\bot\; S_s
\end{equation}

Thus, quantum jumps give rise to the product spacetime
(\ref{1.3.4}) and a particular synchronous reference frame
\begin{equation}
\label{1.3.7} p=(t,s)\leftrightarrow(t,\vec x)
\end{equation}
The latter may be called cosmological reference frame and regarded
as a canonical synchronous reference frame.

In the coordinate-free representation, the metric (\ref{1.3.3})
reads
\begin{equation}
\label{1.3.8} g=dt\otimes dt-h_t,\quad h_t\leftrightarrow
h_{ij}(t,\vec x)dx^idx^j,\quad h_{ij}=-g_{ij}
\end{equation}
in which $h_t$ is a Riemannian metric tensor on $S$ depending on
$t$.

\subsection{Matter}

In what follows, quantities pertaining to matter are: the state
projector $P$ (\ref{1.1.3}), the space part of the energy-momentum
tensor operator
\begin{equation}
\label{1.4.1} T_{{\rm space}}\leftrightarrow T_{ij}
\end{equation}
and the Hamiltonian
\begin{equation}
\label{1.4.2} H_t=\int T_{00}(t,\vec x)d\vec x
\end{equation}
or in the coordinate-free representation
\begin{equation}
\label{1.4.3} H_t=\int \mu(ds)T_{\rm time}(t,s),\quad T_{\rm
time}\leftrightarrow T_{00}
\end{equation}

\section{Energy determinacy}

\subsection{The principle of cosmic energy determinacy}

The cosmic energy determinacy principle reads: The universe is for
all time in a state with a determinate value of energy, i.e., in
an eigenstate of the Hamiltonian $H_t$. Note that a combination of
the energy determinacy and a time-dependent Hamiltonian is
analogous to the adiabatic potential method in quantum scattering
theory.

\subsection{Necessary conditions for the state projector}

A necessary condition for the state projector $P$ that follows
from energy determinacy is
\begin{equation}
\label{2.2.1} H_tP(t)=P(t)H_t=\varepsilon(t)P(t)
\end{equation}
The spectral decomposition of the Hamiltonian is of the form
\begin{equation}
\label{2.2.2} H_t=\sum\limits_l \varepsilon_l(t)P_l(t)
\end{equation}
where $P_l$ is the projector for a level $l$ with an energy
$\varepsilon_l$. For a nondegenerate level
\begin{equation}
\label{2.2.3} {\rm Tr}\;P_l=1
\end{equation}
for a degenerate one
\begin{equation}
\label{2.2.4} {\rm Tr}\;P_l\geq 2
\end{equation}
Let the state belong to a level $l$:
\begin{equation}
\label{2.2.5} P=P^l,\quad {\rm Tr}\;P^l=1
\end{equation}
Then (\ref{2.2.1}) reduces to
\begin{equation}
\label{2.2.6}
P_{l'}P^l=P^lP_{l'}=\delta_{ll'}P^l
\end{equation}
We call $P_l$ a confining projector
\begin{equation}
\label{2.2.7} P_l=P_{\rm con}
\end{equation}

\subsection{Extended confinement}

For what follows, provision should be made for an extension of the
confining projector
\begin{equation}
\label{2.3.1} P_{\rm con}=\sum\limits_{l\in L_{\rm con}}P_l
\end{equation}
where $L_{\rm con}$ is a set of levels. The confinement condition
for $P$ takes the form of
\begin{equation}
\label{2.3.2} P_{\rm con}P=PP_{\rm con}=P
\end{equation}

\section{Equations of motion: Confined dynamics}

\subsection{The state of the universe}

The state of the universe is defined by metric, its time
derivative, and the state projector:
\begin{equation}
\label{3.1.1} \omega=(g,\dot g,P)\Leftrightarrow(h,\dot
h,P)\leftrightarrow(h_{ij},\dot
h_{ij},P)\Leftrightarrow(g_{ij},\dot g_{ij},P)
\end{equation}

\subsection{The equations for metric}

The equations for metric are of the form
\begin{equation}
\label{3.2.1} G_{ij}-\Lambda g_{ij}=8\pi\kappa\tilde T_{ij}
\end{equation}
or in the coordinate-free representation
\begin{equation}
\label{3.2.2} G_{\rm space}+\Lambda h=8\pi\kappa\tilde T_{\rm
space}
\end{equation}
We have
\begin{equation}
\label{3.2.3} G_{\rm space}=G_{\rm space}[g,\dot g,\ddot g],\quad
G_{\rm space}\;{\rm is\; linear\; in}\; \ddot g
\end{equation}
from here on we use square brackets to mean a functional. $T_{\rm
space}$ does not depend on $\ddot g$; $g$ and $\dot g$ are
continuous in time.

\subsection{The equation for the state projector: Confinement}

In the case of
\begin{equation}
\label{3.3.1} {\rm Tr}\;P_{\rm con}\geq 2
\end{equation}
(specifically when $P_{\rm con}=P_l$ and the level $l$ is
degenerate), the confinement condition (\ref{2.3.2}) is
insufficient. An equation of motion for $P$ should be introduced.
 From (\ref{2.3.2}) follows
\begin{equation}
\label{3.3.2} (dP_{\rm con})P+P_{\rm con}dP=(dP)P_{\rm
con}+PdP_{\rm con}=dP
\end{equation}
For any projector $E$
\begin{equation}
\label{3.3.3} E(dE)E=0
\end{equation}
holds. With equations (\ref{2.3.2}) and (\ref{3.3.3}) we can
obtain a solution to (\ref{3.3.2}) for $dP$ of the form
\begin{equation}
\label{3.3.4} dP=PdP_{\rm con}+(dP_{\rm con})P
\end{equation}
Thus the equation of motion for $P$ is
\begin{equation}
\label{3.3.5} \frac{dP}{dt}=\left\{\frac{dP_{\rm
con}}{dt},P\right\}\equiv\frac{dP_{\rm con}}{dt}P+P\frac{dP_{\rm
con}}{dt}
\end{equation}

We assume that the Hamiltonian
\begin{equation}
\label{3.3.6} H=H[g]
\end{equation}
does not depend on $\dot g$. Then in view of (\ref{2.2.2}) and
(\ref{2.3.1})
\begin{equation}
\label{3.3.7} P_{\rm con}=P_{\rm con}[g]
\end{equation}
so that
\begin{equation}
\label{3.3.8} \frac{dP_{\rm con}}{dt}=\frac{dP_{\rm
con}}{dt}[g,\dot g]
\end{equation}
does not depend on $\ddot g$. Thus
\begin{equation}
\label{3.3.9} \frac{dP}{dt}=\left\{\frac{dP_{\rm con}}{dt}[g,\dot
g],P\right\}
\end{equation}
so that $dP/dt$ is given as a functional of the state $\omega$
(\ref{3.1.1}) of the system.

Now
\begin{equation}
\label{3.3.10} \dot H=\dot H[g,\dot g]
\end{equation}
Thus $H$ and $\dot H$ are continuous in time.

\section{Vertices and tentative dynamics}

\subsection{Vertex}

A vertex is a confluence or/and branch point of levels. We assume
that the set of vertices has no accumulation points.

Consider an infinitesimal neighborhood of a vertex. For $t=t_{\rm
vertex}\equiv t_{\rm ver}$ we have
\begin{equation}
\label{4.1.1} H(t_{\rm ver})=H_{\rm ver}(t_{\rm
ver})+H_\perp(t_{\rm ver})
\end{equation}
where
\begin{equation}
\label{4.1.2} H_{\rm ver}(t_{\rm ver})\equiv H_{\rm
ver}=\varepsilon_{\rm ver}P_{\rm ver}
\end{equation}
For $t=t_{\rm ver}\mp 0$ we have
\begin{equation}
\label{4.1.3} H_{\rm ver}(t_{\rm ver}\mp 0)\equiv H_{\rm
ver}^\mp,\quad H_{\rm ver}^+=H_{\rm ver}^-=H_{\rm ver}
\end{equation}
and
\begin{equation}
\label{4.1.4} \dot H_{\rm ver}(t_{\rm ver}\mp 0)\equiv\dot H_{\rm
ver}^\mp,\quad \dot H_{\rm ver}^+=\dot H_{\rm ver}^-=\dot H_{\rm
ver}
\end{equation}
Now
\begin{equation}
\label{4.1.5} H_{\rm ver}^\mp=\sum_l^{1,n^\mp}\varepsilon_l^\mp
P_l^\mp
\end{equation}
where
\begin{equation}
\label{4.1.6} \varepsilon_l^\mp=\varepsilon_{\rm ver}
\end{equation}
\begin{equation}
\label{4.1.7} \sum_l^{1,n^\mp}P_l^\mp\equiv P_{\rm ver}^\mp=P_{\rm
ver}
\end{equation}
Next
\begin{equation}
\label{4.1.8} \dot H_{\rm
ver}^\mp=\sum_l^{1,n^\mp}\dot\varepsilon_l^\mp
P_l^\mp+\sum_l^{1,n^\mp}\varepsilon_l^\mp\dot
P_l^\mp=\sum_l^{1,n^\mp} \dot\varepsilon_l^\mp
P_l^\mp+\varepsilon_{\rm ver}\dot P_{\rm ver}^\mp
\end{equation}
so that
\begin{equation}
\label{4.1.9} \sum_l^{1,n^+}\dot\varepsilon_l^+ P_l^+
+\varepsilon_{\rm ver}\dot P_{\rm
ver}^+=\sum_l^{1,n^-}\dot\varepsilon_l^- P_l^- +\varepsilon_{\rm
ver}\dot P_{\rm ver}^-
\end{equation}
Multiplying both sides of (\ref{4.1.9}) from the left and from the
right by $P_{\rm ver}$ and taking into account (\ref{4.1.7}) we
obtain
\begin{equation}
\label{4.1.10} \sum_l^{1,n^+}\dot\varepsilon_l^+(P_{\rm
ver}P_lP_{\rm ver})^+ +\varepsilon_{\rm ver}(P_{\rm ver}\dot
P_{\rm ver}P_{\rm ver})^+=\sum_l^{1,n^-}\dot\varepsilon_l^-(P_{\rm
ver}P_lP_{\rm ver})^- +\varepsilon_{\rm ver}(P_{\rm ver}\dot
P_{\rm ver}P_{\rm ver})^-
\end{equation}
In view of (\ref{3.3.3}), it follows from (\ref{4.1.10}) and
(\ref{4.1.9}) that
\begin{equation}
\label{4.1.11}
\sum_l^{1,n^+}\dot\varepsilon_l^+P_l^+=\sum_l^{1,n^-}\dot
\varepsilon_l^-P_l^-
\end{equation}
and
\begin{equation}
\label{4.1.12} \dot P_{\rm ver}^+=\dot P_{\rm ver}^- =\dot P_{\rm
ver}
\end{equation}

\subsection{Crossing and tangency}

Consider equation (\ref{4.1.11}). Let
\begin{equation}
\label{4.2.1} \dot\varepsilon_l^+=\dot\varepsilon_l^-\; {\rm
and}\; P_l^+=P_l^-\;\; {\rm for}\; l=1,2,\cdot\cdot\cdot,k,\quad
k<{\rm min}\{n^-,n^+\}\quad {\rm or}\quad k=n^-=n^+
\end{equation}
We call levels $l=1,2,\cdot\cdot\cdot,k$ crossing levels.
For them
\begin{equation}
\label{4.2.2} \sum_l^{1,k}P_l^+=\sum_l^{1,k}P_l^-\equiv P_{\rm
cross}
\end{equation}
Now
\begin{equation}
\label{4.2.3}
\sum_l^{k+1,n^+}\dot\varepsilon_l^+P_l^+=\sum_l^{k+1,n^-}\dot
\varepsilon_l^-P_l^-
\end{equation}
and
\begin{equation}
\label{4.2.4} \sum_l^{k+1,n^+}P_l^+ =\sum_l^{k+1,n^-}P_l^-\equiv
P_{\rm tan}
\end{equation}
where, in view of what will follow, `tan' stands for tangency, and
\begin{equation}
\label{4.2.5} P_{\rm cross}+P_{\rm tan}=P_{\rm ver}
\end{equation}
Let $k+1<{\rm min}\{n^-,n^+\}$ and, for the sake of definiteness,
$n^+\geq n^-$. From (\ref {4.2.3}) follows
\begin{equation}
\label{4.2.6}
P_{l'}^+\sum_l^{k+1,n-}\dot\varepsilon_l^-P_l^-P_{l''}^+=0,\quad
l',l''\in\{k+1,\cdot\cdot\cdot,n^+\},\;\;l''\neq l'
\end{equation}
Substituting
\begin{equation}
\label{4.2.7} P_{k+1}^-=P_{\rm tan}-\sum_l^{k+2,n^-}
\end{equation}
into (\ref{4.2.6}) we obtain
\begin{equation}
\label{4.2.8} \sum_l^{k+2,n^-}(\dot\varepsilon_l^-
-\dot\varepsilon_{k+1}^-)P_{l'}^+P_l^-P_{l''}^+=0,\quad k+2\leq
n^-
\end{equation}
The number of the quantities $(\dot\varepsilon_l^-
-\dot\varepsilon_{k+1}^-)$ is $[n^- -(k+1)]$, the number of
equations is no less than
\begin{equation}
\label{4.2.9} \frac{(n^- -k)^2-(n^- -k)}2=\frac{n^- -k}2[n^-
-(k+1)]\geq n^- -(k+1)
\end{equation}
so that
\begin{equation}
\label{4.2.10} \dot\varepsilon_l^-=\dot\varepsilon_{k+1}^-,\quad
l=k+2,\cdot\cdot\cdot,n^-
\end{equation}
Now
\begin{equation}
\label{4.2.11}
\sum_l^{k+1,n^+}\dot\varepsilon_l^+P_l^+=\dot\varepsilon_{k+1}^-P_{\rm
tan}
\end{equation}
whence
\begin{equation}
\label{4.2.12} \dot\varepsilon_l^+=\dot\varepsilon_{k+1}^-,\quad
l=k+1,\cdot\cdot\cdot,n^+
\end{equation}
Thus
\begin{equation}
\label{4.2.13} \dot\varepsilon_{k+1}^+=\cdot\cdot\cdot=
\dot\varepsilon_{n^+}^+=\dot\varepsilon_{k+1}^-
=\cdot\cdot\cdot=\dot\varepsilon_{n^-}^-\equiv\dot\varepsilon_{\rm
tan}
\end{equation}
which means tangency of the levels $l\geq k+1$.

\subsection{Tentative vertex dynamics}

Let at $t=t_{\rm ver}-0$ the state belong to a level $\bar
l\in\{1,2,\cdot\cdot\cdot,n^-\}$:
\begin{equation}
\label{4.3.1} P(t_{\rm ver}-0)=P^{\bar l}(t_{\rm ver}-0)
\end{equation}
Confined dynamics along with energy determinacy imply equations
\begin{equation}
\label{4.3.2} P_{\rm con}(t_{\rm ver}-0)=P_{\bar l}(t_{\rm
ver}-0),\qquad P_{\rm con}(t_{\rm ver}+0)=P_{\rm con}(t_{\rm
ver})= P_{\rm ver}
\end{equation}
That is the tentative vertex dynamics.

\section{Quantum jumps and their probabilities:\\ Actual dynamics
}
\subsection{Energy determinacy and jumps}

In the case of branching $(n^+\geq 2)$, the tentative vertex
dynamics would violate energy determinacy. To prevent the
violation, quantum jumps must be introduced. Thus we have
\begin{equation}
\begin{array}{l}
\label{5.1.1}
   P(t_{\rm
ver}-0)=P^{\bar l}(t_{\rm ver}-0)\stackrel{\rm
jump}{\longrightarrow}P(t_{\rm ver})=P^l(t_{\rm ver}+0)=
\frac{\displaystyle
  P_l^+P^{\bar l}(t_{\rm ver}-0)P_l^+}
  {\displaystyle
   {\rm
Tr}\{P_l^+P^{\bar l}(t_{\rm ver}-0)\}}\; ,\\
\\
  \qquad\qquad\qquad {} \bar
l\in\{1,2,\cdot\cdot\cdot,n^-\},\quad
l\in\{1,2,\cdot\cdot\cdot,n^+\},\quad n^+\geq 2
\end{array}
\end{equation}
The state is continuous on the right.
For the confining projector,
we obtain
\begin{equation}
\label{5.1.2} P_{\rm con}(t_{\rm ver}-0)=P_{\bar l}(t_{\rm
ver}-0)\stackrel{\rm jump}{\longrightarrow}P_{\rm con}(t_{\rm
ver})=P_l(t_{\rm ver}+0)
\end{equation}
The set
\begin{equation}
\label{5.1.3} \{P_l^+\}_{l=1}^{n^+}
\end{equation}
is determined by
\begin{equation}
\label{5.1.4} H(t_{\rm ver}+\tau)=H[g(t_{\rm ver}+\tau)]\quad {\rm
for}\; \tau\rightarrow 0
\end{equation}
i.e., by
\begin{equation}
\label{5.1.5} H[g(t_{\rm ver})+\dot g(t_{\rm ver})\tau]\quad {\rm
for}\; \tau\rightarrow 0
\end{equation}
which is independent of $\ddot g(t_{\rm ver})$ and, therefore, of
$P(t_{\rm ver}-0)$.

\subsection{Probabilities}

The probability of the jump $\bar l\rightarrow l$ (\ref{5.1.1}) is
\begin{equation}
\label{5.2.1} w^{\bar l\rightarrow l}={\rm Tr}\{P_l^+P^{\bar
l}(t_{\rm ver}-0)\}
\end{equation}

In the case of crossing there is no jump.

\subsection{Actual dynamics}

The actual dynamics is the level-confined one:
\begin{equation}
\label{5.3.1} P_{\rm con}=P_l
\end{equation}
It involves quantum jumps. They are a reaction against the
propensity of the universe dynamics to the violation of energy
determinacy.

\section*{Acknowledgments}

I would like to thank Alex A. Lisyansky for support and Stefan V.
Mashkevich for helpful discussions.


\begin{thebibliography}{9}

\bibitem{1} W. Heisenberg, The Physical Principles
of the Quantum Theory
(University of Chicago Press, 1930).
\bibitem{2} Feynman Lectures on Gravitation
(Addison-Wesley Publishing Company, 1995)
\bibitem{3} Vladimir S. Mashkevich, gr-qc/9409010 (1994).
\bibitem{4} Vladimir S. Mashkevich, gr-qc/9505034 (1995).
\bibitem{5} Vladimir S. Mashkevich, gr-qc/0103051 (2001).
\bibitem{6} Steven Weinberg, Gravitation and Cosmology
(Wiley, New York, 1973).
\bibitem{7} L.D. Landau, E.M. Lifshitz,
The Classical Theory of
Fields (Pergamon Press, Oxford, 1975).
\bibitem{8} Malcolm Ludvigsen, General Relativity
(Cambridge University Press, 1999).
\bibitem{9} J.A. Peacock, Cosmological Physics (Cambridge
University Press, 1999).

\end{thebibliography}
\end{document}